\documentclass[useAMS,usenatbib,usegraphicx]{mn2e}

\begin{document}

\title { An efficient PSF construction method } 

\author { M. Gai \thanks{ INAF - Astron. Observatory of Turin }, 
R. Cancelliere \thanks{ Dep. of Comp. Sciences, Univ. of Turin }} 

\author[M. Gai and R. Cancelliere]{M. Gai$^{1}$ 
\thanks { E-mail: gai@oato.inaf.it (MG), cancelli@di.unito.it (RC)} 
and R. Cancelliere$^{2}$ \\ 
$^{1}$Istituto Nazionale di Astrofisica - Osservatorio Astronomico di 
Torino, V. Osservatorio 20, 10025 Pino T.se (TO), Italy \\ 
$^{2}$Dipartimento di Informatica, Universit\`a di Torino, 
C.so Svizzera 185, 10149 Torino, Italy }

\date{Accepted 2007 March 1. Received 2007 March 1; in original form 
2006 December 18 }

\pagerange{\pageref{firstpage}--\pageref{lastpage}} \pubyear{2007}

\maketitle

\label{firstpage}

\begin{abstract}
Image computation is a fundamental tool for performance assessment 
of astronomical instrumentation, usually implemented by Fourier transform 
techniques. 
We review the numerical implementation, evaluating a direct implementation 
of the discrete Fourier transform (DFT) algorithm, compared with fast 
Fourier transform (FFT) tools. 
Simulations show that the precision is quite comparable, but in the case 
investigated the computing performance is considerably higher for DFT than 
FFT. 
The application to image simulation for the mission Gaia and for Extremely 
Large Telescopes is discussed. 
\end{abstract} 


\section{Introduction}
\label{intro}
The imaging performance of astronomical instruments, and optical 
instrumentation in general, is related to the Point Spread Function 
(PSF), i.e. the intensity distribution from a point-like object at 
infinity, which is in most cases a satisfactory representation of 
a star. 
The monochromatic PSF is computed as square modulus of the field amplitude 
on the focal plane; the amplitude function, in turn, is built through the 
diffraction integral, derived from the wave description of electromagnetic 
radiation. 
An historical review of the diffraction theory is out of the scope of 
this paper, devoted to performance comparison of two Fourier transform 
(FT) algorithms computing the diffraction integral. 

The PSF includes, by a suitable wavefront description, details of 
the instrumental response. 
The polychromatic PSF of an unresolved star is produced by 
superposition, weighed by spectral distribution, of the monochromatic 
PSFs; thus, objects with different spectral characteristics have 
different polychromatic PSF. 
The image of a resolved source can be described as a superposition 
of PSFs, weighed by spectrum and spatial distribution. 
The response of a spectrograph can be built in quite the same way, 
apart the explicit wavelength dependence of the monochromatic PSF 
location. 
\\ 
PSF construction is therefore a quite general tool. 

An accurate PSF description for an optical instrument, either in design 
or observation phase (through calibration), is crucial to any 
astrophysical measurement. 
In the former case, system parameters are tuned to improve the 
performance, whereas in the latter the instrument signature must be 
identified for measurements correction and determination of the 
intrinsic objects characteristics. 
\\ 
The PSF is usually computed numerically, although useful 
analytical approximations are derived e.g. in \citet{braat}. 
The FFT is used for PSF construction also by 
professional ray tracing packages (e.g. Zemax or Code~V). 
\\ 
We discuss the numerical implementation of the FT, through a method 
of direct computation which, under appropriate conditions, may be 
more convenient than the usual Fast Fourier Transform (FFT) tools. 
The computing time required by the proposed PSF construction method is 
significantly lower than the equivalent FFT case; also, the result 
precision can be higher, since the generation of some artefacts is 
suppressed. 
Both aspects are relevant to applications, with benefits on processing, 
resolution and/or precision. 
\\ 
In most of our simulations, the case study is the telescope of the Gaia 
mission \citep{Gaia}, approved by the European Space Agency for launch in 
2011, and aimed at a high precision astrometric survey of our Galaxy. 
The science data simulations require high accuracy, for proper validation 
of the reduction algorithms under development by the astrophysical 
community involved. 
A specific application to wavelength dependence, already dealt with by 
the authors \citep{gai05}, is further developed in \citet{canc07}, 
recently submitted. 
We also mention the possible application of the proposed PSF construction 
method to the case of Extremely Large Telescopes, which among 
other technological challenges also feature impressive numerical 
requirements. 
\\ 
In section \ref{sec_diffraction} we recall the basics of the diffraction 
integral and discuss its key implementation features. 
In section \ref{sec_FFT_DFT} we compare some of the more relevant features 
of FFT and DFT implementations, in terms of processing time and accuracy. 
Then, in section \ref{sec_spectral}, we discuss the PSF dependence from 
wavelength and the possible introduction of artefacts. 
In section \ref{sec_OWL} we assess the application of the proposed method 
to modern Extremely Large Telescopes. 
Finally, we draw our conclusions.

\section { The diffraction integral } 
\label{sec_diffraction} 
The definitions related to imaging performance derive from a wide range 
of applications; hereafter, we mostly follow the notation from \citet{born}. 
\\ 
A point-like source at infinity, as can be considered a star except for 
very high spatial resolution instruments (like interferometers), generates 
a spherical wavefront which can be considered flat at the entrance of a 
telescope. 
This description neglects the disturbances in the medium, e.g. air 
turbulence; in modern ground based telescopes, an Adaptive Optics (AO) 
system is used to recover, at least in part, the ideal diffraction 
limited performance. 

The telescope, ideally, folds the input wavefront towards a single point 
of the focal plane (FP), where the diffraction image is achieved. 
The imaging performance of a real optical system can usually be described 
by the discrepancy with respect to ideal propagation, mathematically 
represented by an equivalent deviation from the input flat wavefront, i.e. 
the wavefront error (WFE). 
\\ 
The diffraction image from a real telescope is built by the square 
modulus of the diffraction integral, here represented in circular 
coordinates $\left\{r,\ \phi\right\}$ and $\left\{\rho,\ \theta\right\}$, 
respectively on FP and pupil: 
\begin{equation}
\label{eq:PSF_circ}
I\left( r,\phi \right) = 
k \left| \int \rho \, d\rho \, d\theta \, 
P \left( \rho ,\theta \right) e^{-i\pi r\rho \,\cos \left(
\theta -\phi \right) } \right| ^{2} \, . 
\end{equation}
The integration domain corresponds to the pupil; for a circular aperture, 
$0 \le \rho \le 1; \, 0 \le \theta \le 2\pi$, where $\rho$ is the 
normalised radius. 
The constant $k$ provides the appropriate photometric result associated 
to the source emission, exposure time and collecting area; other scaling factors 
are neglected. 
The pupil function 
$P \left( \rho ,\theta \right) = e^{i \Phi \left( \rho ,\theta \right)}$, 
depends on the phase aberration function $\Phi$, which is usually expanded 
in a series of terms, e.g. the Zernike functions $\phi_n$ \citep{born}: 
\begin{equation}
\label{eq:aberr}                                                                                       
\Phi \left( \rho ,\theta \right) = \frac{2\pi}{\lambda} WFE = 
\frac{2\pi}{\lambda} \sum_n  A_n \phi_n 
\left( \rho, \theta \right) \ . 
\end{equation} 
The WFE itself is independent from wavelength, contrarily to the pupil 
function, which is affected by the $2 \pi / \lambda$ factor. 
Also, the nonlinear relationship between the set of aberration coefficients 
$A_n$ and the image is put in evidence by replacement of Eq. (\ref{eq:aberr}) 
in Eqs. \ref{eq:PSF_circ} or \ref{eq:PSF_rect}. 

The non aberrated case, corresponding to $WFE = 0$, can be solved, 
providing an analytic representation of the diffraction image of simple 
systems recalled in appendix \ref{app_diffraction}. 
\\ 
The form of the diffraction integral, used in Eqs. \ref{eq:PSF_circ} 
and \ref{eq:PSF_rect}, leads naturally to its implementation based on 
FT techniques.

\subsection { Discrete and Fast FT } 
\label{sec_DFT} 
The Discrete Fourier Transform (DFT), e.g. from \citet{bracewell}, 
of a one-dimensional sequence $s_n = s(z_n)$ of $N$ samples of the signal 
$s$, function of a time variable $z$, is: 
\begin{equation}
\label{eq:fourier} 
S(\omega) = \sum_{n=0}^{N-1} s_n \exp(-i z_n \omega) \, , 
\end{equation}
where $\omega$ is the pulsation, conjugated to the signal 
argument. 
The signal $s$ is supposed to be smooth and with finite duration, with 
$N$ contiguous non-zero samples; this ensures proper convergence of 
Eq. \ref{eq:fourier}. 
$S(\omega)$ is evaluated on a set of $M$ points in the desired 
interval. 
Apart variable transformation, the diffraction integral in Eq. 
\ref{eq:PSF_circ} actually appears as the FT of the pupil function, 
with the natural extension to two dimensions of 
Eq. \ref{eq:fourier}, and the discrete numerical form required 
for practical computation cases. 
Some remarks on sampling resolution and processing time are provided 
in \ref{app_FT}. 

In the Fast FT (FFT) algorithm, significant improvements in the 
computation time are achieved under the restriction (usually quite 
acceptable) that the signal is uniformly sampled, so that $s_n = s(nT)$, 
with sampling period $T$. 
The computation of all $N$ transform values, for each $\omega_q$, 
is performed as a single process, split in a hierarchical sequence of 
lower order FFT steps. 

The diffraction integral is physically limited to the real pupil size; 
however, with this limitation, only two points are generated over the 
characteristic length of the system, which is clearly not sufficient 
to provide an acceptable image detail. 
A typical solution consists in a formal extension of the pupil domain to 
larger size, allowing to achieve the desired resolution, provided the 
pupil function $P$ is set to zero in the external region. 
In practice, the real pupil must be extended to a dummy pupil $M$ times 
larger, in order to force the computation of $M$ points in the Airy 
radius. 
The idea of zero padding for specific FT computations 
has been applied e.g. to the packing theorem \citep{bracewell}; 
it is a common approach also in ray tracing packages (Zemax, Code~V). 

The pupil resolution must be sufficient to sample the intrinsic 
variation of the WFE, e.g. placing $Q$ points over the pupil size. 
This generates an array of $N = M \times Q$ points mapping the dummy pupil 
along one dimension; notably, $(M-1) Q$ of them have zero value. 
This array can be mapped by DFT onto $M \times Q$ points over the FP, and 
the transform array covers $Q$ times the Airy radius. 

\subsection { PSF construction by FT }
In the PSF construction according to the above prescription, the FFT 
approach involves a significant amount of computation over quantities 
on pupil set to zero. 
Also, the FP region of interest (ROI) is usually not as large as 
implicitly computed by the FFT, because after a few Airy diameters 
the PSF becomes vanishingly small for most optical systems. 
This consideration led us to the alternative approach of direct 
DFT computation according to Eq. \ref{eq:fourier} \citep{canc07}. 
The key features of this strategy are: 
\\ 
-- restriction to the physical pupil only; 
\\ 
-- restriction to a limited FP region. 
\\ 
Thus, the case considered concerns the comparison between FFT applied 
to a square image of logical format $N \times N$, where $N = M \times Q$, 
and DFT applied to an input array $Q \times Q$ to generate an array 
$M \times M$. 
For convenience of some numerical examples below, we will set $M = Q$, 
although this is by no means a restriction to the method.

\section { Test of FFT and DFT }
\label{sec_FFT_DFT}
The first relevant comparison between FFT and DFT algorithms concerns their 
effectiveness, i.e. their capability of providing the correct result. 
Both are derived from the same diffraction integral, so that, apart the 
case of actual coding errors, they both rely on the same convergence 
conditions, and are expected to provide the same result, within numerical 
errors. 
The simulation is implemented on a desktop computer with a Pentium II 
processor, clock 2~GHz, 512 MB RAM, in the Matlab environment 
(http://www.mathworks.com). 
The Matlab {\em fft} function is based on the FFTW library 
(http://www.fftw.org), described in \citet{fftw}, which is commonly considered 
to be quite efficient, whereas the DFT is implemented by straightforward 
coding of Eq. \ref{eq:PSF_circ} for the rectangular aperture case, as in 
appendix \ref{app_diffraction}. 

\subsection { Non aberrated PSF } 
\label{sec_ideal_PSF}
The first test is performed on non aberrated images, since this is the only 
practical case in which an analytic representation of the diffraction image 
is readily available (appendix \ref{app_diffraction}). 
The Gaia telescope parameters are used: rectangular aperture, size 
$D_x = 0.5$~m times $D_y = 1.4$~m (high resolution in the $y$ direction); 
focal length $F = 35$~m. 
The ROI on the FP is restricted to $\pm 600~\umu$m ($x$) $\times \pm 
200~\umu$m ($y$), with resolution respectively 6 and 2~$\umu$m, and the 
DFT is computed only over this region, using the FFT sampling points 
over both pupil and FP. 

The DFT or FFT precision improves with increasing resolution of the 
sampled function, i.e. with increasing number of points. 
The PSF discrepancy between the FFT and DFT result, respectively, and 
the analytic expression, is computed vs. increasing resolution in the 
pupil plane, over the range $Q = 10$ to 64 sampling points. 
Since the actual FFT image format is $Q^4$, the largest array generated 
is $4K \times 4K$, i.e. 16 million pixels. 
At this point, our desktop computers already have a significant virtual 
memory usage. 
\begin{figure}
\includegraphics[width=80mm]{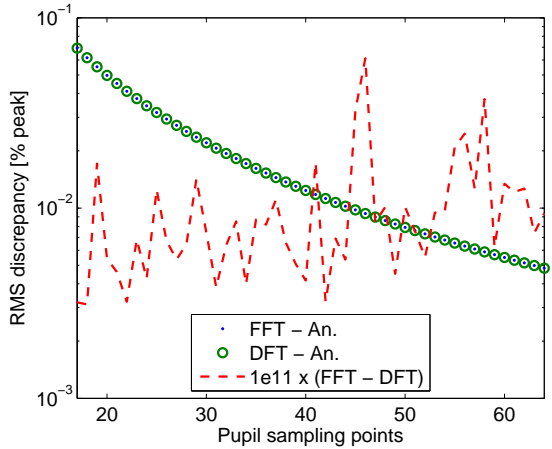}
\caption { RMS discrepancy between FFT, DFT and analytic non aberrated 
PSF vs. number of sampling points. } 
\label{fig_RMS_FDA}
\end{figure}
\begin{figure}
\includegraphics[width=80mm]{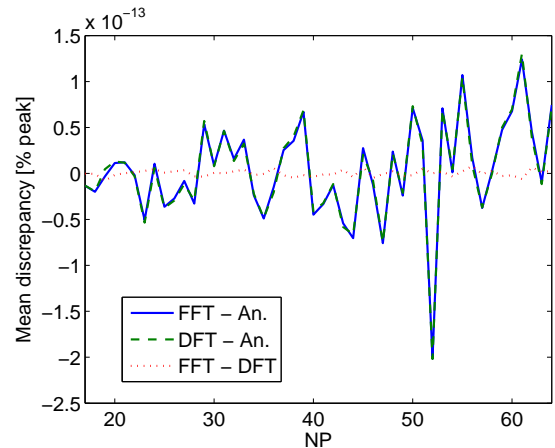}
\caption { Mean discrepancy between FFT, DFT and analytic non aberrated 
PSF vs. number of sampling points. } 
\label{fig_mean_FDA}
\end{figure}
The discrepancy is evaluated as mean, peak to valley (hereafter ``peak´´ 
for simplicity) and RMS value of the difference between computed PSFs, 
evaluated on the whole ROI. 
The RMS discrepancy of both FFT and DFT with respect to the analytic PSF 
(shown in Fig. \ref{fig_RMS_FDA}, respectively by dots and circles) 
decreases with increasing sampling resolution, as expected; also, 
the RMS difference between FFT and DFT (dashed line) is quite small, 
$\sim 1e-15$ of the PSF peak. 
Since the minimum real number used (the Matlab constant {\em ``eps"}) 
is $2.22e-16$, the discrepancy between DFT and FFT results is quite 
close to the limiting numerical noise. 
Thus, both FFT and DFT compute the desired approximation to the analytic 
PSF, with negligible difference between them. 
\\ 
A similar improvement with sampling resolution is shown by the peak 
discrepancy of FFT and DFT vs. analytic PSF; also, the mean discrepancy 
remains quite small (order of $1e-15$ of the PSF peak value) over the 
whole range, as shown in Fig. \ref{fig_mean_FDA}. 
In particular, when the pupil is sampled with more than $14 \times 14$ 
points the RMS discrepancy over the ROI drops below 0.1\% of the peak 
value, and the $1e-4$ level is reached for more than $45 \times 45$ 
points in the pupil. 
Similarly, the peak discrepancy is below 1\% when using more than 
$14 \times 14$ points and below 0.1\% with sampling above $48 \times 48$ 
points.

\subsection { Aberrated PSF } 
\label{sec_gen_PSF}
The non aberrated PSF case, although quite useful for comparison of 
the two FT methods, is not particularly interesting from an optical 
standpoint, since any real system has deviations from the ideal case. 
However, for a general aberrated case, an analytic representation is 
{\em not} available, so that the above simulation cannot be extended 
to the images related to an arbitrary case. 
However, we can build a set of aberrated cases and evaluate the 
PSF discrepancy between DFT and FFT results, which can be compared with 
the corresponding non aberrated case. 
\\ 
We generate a set of 1000 aberration cases, with sampling resolution of 
24 points, using a Gaussian distribution of the Zernike coefficients 
with $\sigma = 100$~nm. 
\begin{figure}
\includegraphics[width=80mm]{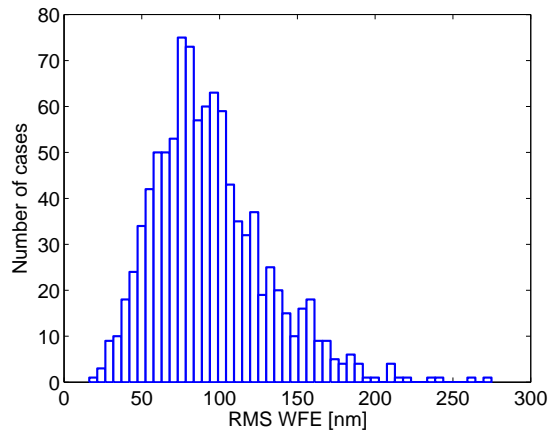}
\caption { Histogram of the RMS WFE } 
\label{fig_wfe}
\end{figure}
The resulting RMS WFE distribution is shown in Fig. \ref{fig_wfe}. 
The average value is 94~nm, i.e. an acceptable value for realistic 
optics in the visible ($\sim \lambda / 6$ at $\lambda = 600$~nm), 
with standard deviation 36~nm. 

For each set of aberration coefficients, the PSF is built by both FFT 
and DFT, then the FFT result is cut to the size of the selected FP ROI, 
where also the DFT is computed. 
The discrepancy between the two PSF versions is evaluated, deriving for 
each case the mean, peak and RMS values over the ROI, and recording the 
values for the whole the statistical sample of aberration cases. 

The results are shown in table \ref{tab_discr}. 
The PSF difference between the FFT and DFT methods, for a general case 
of aberration, remains quite small, comparable to their discrepancy for 
the non aberrated case ($\sim 1e-15$ of the peak), and compatible with 
the numerical noise already experienced in the previous test. 
Therefore, also for the general case of an arbitrary PSF, the DFT method 
generates a very good representation of the FFT result. 

\begin{table} 
\caption { Statistics of mean, peak and RMS discrepancy between DFT 
and FFT result, over the sample of 1000 aberration cases. }
\begin{center}       
\begin{tabular}{lcc} 
\hline
		& Sample Mean & Sample RMS \\ 
\hline
ROI mean discrepancy & $5.7e-19$ & $2.8e-17$ \\ 
ROI peak  discrepancy & $3.3e-15$ & $1.9e-15$ \\ 
ROI RMS  discrepancy & $1.5e-16$ & $1.1e-16$ \\ 
\hline
\end{tabular} 
\label{tab_discr}
\end{center}
\end{table}

\subsection { Processing time } 
\label{sec_proc_time} 
The processing time for a one-dimensional FFT on $N$ points is 
order of $N \log N$, and $\sim N^2$ for DFT. 
We deal with two-dimensional images of logical format $N \times N = 
Q^2 \times Q^2$, in which the region of interest is of order of 
$Q \times Q$ pixels. 
The FT computational load can be explicitly derived for the 
bi-dimensional case corresponding to PSF computation. 
\\ 
We consider the case in which the one-dimensional FT algorithm is 
repeated $N = Q^2$ times for the FFT, and $Q$ times for the DFT, 
although this is not necessarily the most efficient implementation. 
Then, the cost for bi-dimensional FFT and DFT becomes of order of 
$N^2 \log N^2 = Q^4 \log Q^4 \propto Q^4 \log{Q}$, 
and $Q^2 \times Q = Q^3$, respectively. 
Introducing additional numerical factors (e.g. the Airy diameters in 
the ROI), the scaling law with respect to the number of sampling points 
$Q$ does not change, apart the coefficient. 
\\ 
The potential gain of DFT vs. FFT is thus $\sim Q \log Q$. 

Notably, similar gain may be expected in other applications in which the 
input data or the FT results are required with high resolution only over 
a restricted region for both conjugate variables. 
We remark also that the condition of uniform sampling is not necessarily 
true for any measurement, and can be sometimes inconvenient. 
The above estimate of processing time is valid only within the 
computation-limited regime, and for any real computer, at increasing 
size of the processed arrays, the processing becomes input/output-limited, 
when the physical memory is saturated and the virtual memory mechanisms 
start swapping data towards the mass storage devices. 
\begin{figure}
\includegraphics[width=80mm]{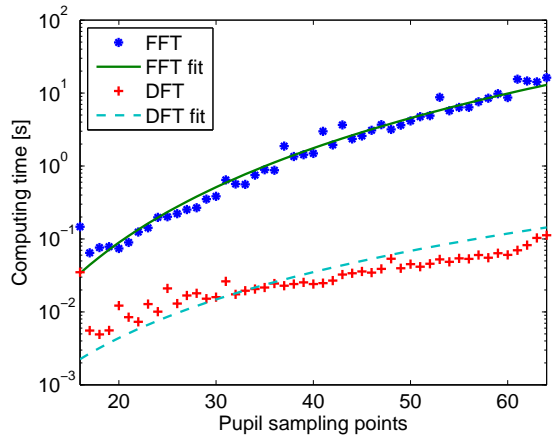}
\caption { Processing time vs. sampling points of FFT (stars) and DFT 
(crosses), and corresponding fitting functions. } 
\label{fig_proc_time}
\end{figure}

A stopwatch timer function of Matlab is used to keep track of the 
execution time of both FFT and DFT. 
The simulation is repeated for 10 iterations, in order to reduce 
the fluctuations due to operating system overheads. 
The processing time is averaged over the 10 sets, 
providing the results shown in Fig. \ref{fig_proc_time}, where the 
fit to the analytic functions $Q^4 \log Q$ (FFT) and $Q^3$ 
(DFT), is also shown. 
The fit to experimental data is reasonable, although noisy; the fit 
parameters are respectively 129~ns (FFT) and 548~ns (DFT). 

Over the range of cases considered, DFT is significantly more 
efficient than FFT, and the growth rate of the computational 
cost is lower.

\section{ Wavelength dependence } 
\label{sec_spectral} 
Since in our examples the FP resolution is fixed (2~$\umu$m), the pupil resolution scales with the wavelength $\lambda$: the pupil is better 
sampled at shorter wavelength and vice versa. 
The variable pupil resolution is potentially associated to artefacts 
in the images, depending on wavelength, leading to potential measurement 
errors, as in \citet{bus06}. 
\\ 
In order to test this aspect, the wavelength range $\lambda_1 = 590$~nm 
to $\lambda_2 = 610$~nm is explored with $\Delta\lambda = 1$~nm resolution. 
The case considered is the non-aberrated optical PSF, because of the 
known analytic representation. 
We evaluate the discrepancy with respect to the analytical function of 
three different PSF representations, respectively built one by FFT, and 
two by DFT, the former using the FFT pupil sampling resolution (DFT~1), 
and the latter using fixed sampling of 64 points on the pupil (DFT~2). 
Notably, DFT~2 resolution is lower than FFT or DFT~1, using respectively 
136 ($\eta$) and 146 ($\xi$) points at $\lambda = 600$~nm. 

\begin{figure}
\includegraphics[width=80mm]{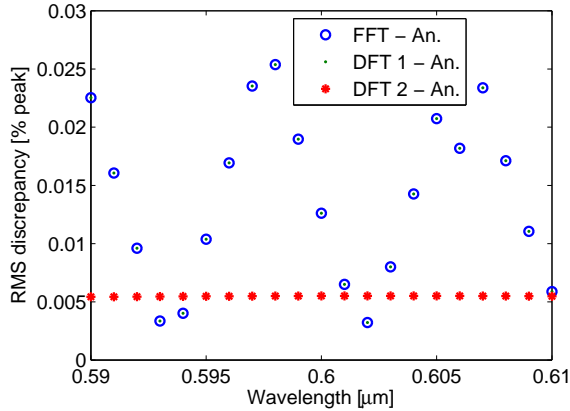}
\caption { PSF discrepancy vs. wavelength of FFT (circles), DFT~1 (dots) 
and DFT~2 (stars). } 
\label{fig_discr_lambda_fft_dft}
\end{figure}

The variation with wavelength of the three cases of PSF discrepancy 
with respect to the analytic representation is shown in Fig. 
\ref{fig_discr_lambda_fft_dft}, respectively for the results from FFT 
(circles), DFT~1 (dots), and DFT~2 (stars). 
The DFT~1 resolution was chosen to match the FFT resolution, so that 
it is reasonable to expect similar behaviour. 
We remark the large variation in discrepancy of FFT and DFT~1 over a 
comparably narrow spectral range. 
This is expected to lead to potentially significant errors in the 
construction of polychromatic images, {\em depending on the selected 
spectral sampling}. 
\\ 
Besides, using a constant pupil sampling vs. wavelength (as for DFT~2, 
compatible with fixed FP resolution), we get a uniform distribution of 
PSF discrepancy vs. wavelength. 
In spite of lower pupil resolution, the precision of DFT~2 is 
significantly better over most of the wavelength range, and above all 
it is constant, i.e. wavelength independent. 

\section{ Extremely Large Telescopes } 
\label{sec_OWL} 
The next generation of telescopes currently studied aim at a significant 
improvement in sensitivity with respect to current instrumentation, as for 
the OverWhelmingly Large telescope (OWL), in \citet{gilmozzi}. 
Of course, the related cost and engineering challenge is also significant. 
Other proposed Extremely Large Telescopes (ELTs) also have typical 
size of a few ten meters, sometimes achieved as diluted apertures, as 
proposed e.g. in \citet{20_20}, for a better trade-off between complexity 
and sensitivity. 

The PSF computation for such instruments is in principle based on the 
same tools (diffraction integral) used for previous telescopes. 
Besides, processing improvements may be appealing to the purpose of 
increased precision in the analysis and more efficient design 
development. 

In recent propositions, the OWL diameter is 42~m; as an example, 
we assume operation in the K band (2.2~$\umu$m), corresponding to an 
Airy diameter of 26~milli-arcsec, and requiring an effective focal 
length of 560~m for its imaging on four 18~$\umu$m pixels. 
For simulation of the OWL PSF, we assume a minimum sampling of order of 
10 points per pixel, i.e. 40 points over the Airy diameter, corresponding 
to 1.8~$\umu$m resolution on the focal plane. 
We set again the ROI size on the FP to 6 Airy diameters (on each dimension), 
for a logical format of $240 \times 240$ pixels. 

For proper sampling of the atmospheric turbulence over the telescope pupil, 
in K band, we set a resolution of order of 0.1~m, requiring an array of 
$420 \times 420$ points on pupil. 
However, the FFT size associated to the problem, due to the compound 
effect of 40 points in the Airy diameter times 420 points on the pupil, 
is $16800 \times 16800$, simply unmanageable on the desktop 
class computer used in our simulation. 

\begin{figure}
\includegraphics[width=80mm]{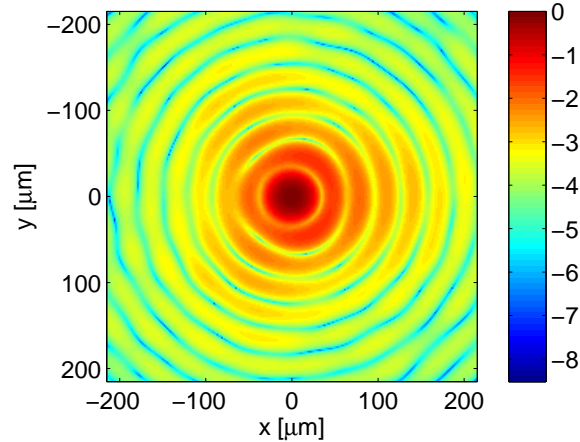}
\caption { Example of an OWL PSF with static aberrations and AO 
residuals. } 
\label{fig_OWL_AO}
\end{figure}

The simulation is set with a few Zernike coefficients set to non zero 
values, to mimic a realistic optical response, and with a Gaussian 
random residual from AO correction, with $\sigma = 300$~nm. 
The corresponding RMS WFE on the pupil in the case shown is $310$~nm. 
The resulting PSF is shown in Fig. \ref{fig_OWL_AO}, in logarithmic 
units; the signal profile is affected by perturbations both on low 
and high frequency, respectively from optical aberrations and AO 
residuals. 
The monochromatic PSF computation requires about 0.6~s, increasing to 
about 5~s for 0.05~m pupil resolution. 
Therefore, analyses of such optical system, either with respect to 
instrumental aspects (including AO response) or to astrophysical 
performance, can be done without high-end computers. 
Besides, for a given computing power, the proposed method 
allows a significant increase of the range of study cases, e.g. for 
statistical evaluation.

\section{Conclusion}
\label{Conclusions}
When PSF computation is required only on a limited region of the focal 
plane, DFT computation can be significantly more efficient than FFT, 
providing the same results (within the numerical noise) with much lower 
processing time. 
This is because computation is restricted to the actual points required 
by the problem, rather than those imposed by the FFT definition on both 
input and output variables. 
Both FFT and DFT produce a good PSF representation, with precision 
increasing with the number of sampling points used. 
Artefacts can be introduced in the PSF, e.g. as a function of wavelength, 
depending on the sampling strategy; DFT provides more flexibility 
and can thus improve the precision. 

Processing time is relevant also with respect to precision of the PSF 
computation and of the performance analysis based on such data; in 
particular, precision is in many cases dependent on the number of 
sampling points used for representation of the relevant problem 
variables. 
Therefore, the capability of processing higher resolution data, in a 
given time frame, ensures higher precision on each PSF; besides, it is 
also possible to take advantage of the proposed method to improve the 
statistical sampling of the problem domain, by generation of a 
larger amount of simulated data, depending on the most convenient 
trade-off. 

Also, the development of new generation, large size 
telescopes may benefit from faster and more accurate computation 
tools, for both design and science assessment.

\section*{Acknowledgment}
The authors would like to thank M.G. Lattanzi, D. Busonero, and 
D. Loreggia for useful comments on the possible application of 
the proposed methods. 
The paper readability benefits of the referee's remarks.


\appendix 

\section{ Ideal diffraction images }
\label{app_diffraction} 

In the simple case of an unobstructed circular pupil of diameter $D$, 
at wavelength 
$\lambda$, and with focal length $F$, we get the radial symmetric PSF 
of Eq. \ref{eq:airy}: 
\begin{equation} 
\label{eq:airy} 
I\left( r\right) = k \left[2 {{J_{1} \left( \pi r D / \lambda F\right) } 
\over {\pi r D / \lambda F }} \right] ^{2} \, .  
\end{equation}
Here $J_1$ is the Bessel function of the first kind, order one, and 
$r$ the radial FP coordinate. 
The image characteristic size is the Airy diameter: $2.44 \lambda F / D$. 

\begin{figure}
\includegraphics[width=80mm]{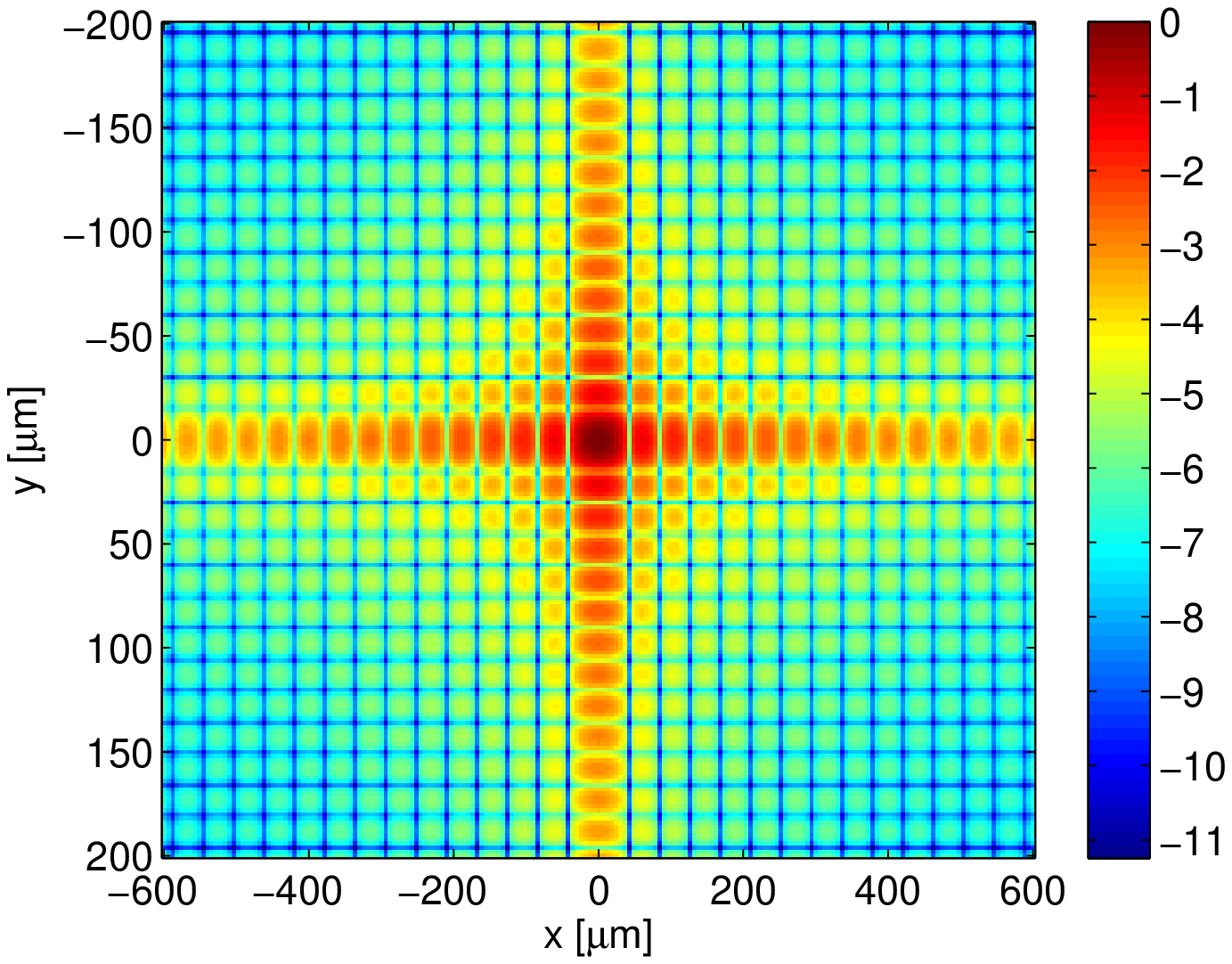}
\caption { Non aberrated PSF of Gaia: logarithmic intensity at 600~nm. } 
\label{fig_PSF_Gaia}
\end{figure}

For a rectangular pupil, it is convenient to use cartesian coordinates, 
e.g. $\left\{ x, \, y \right\}$ on FP and $\left\{ \xi, \, \eta 
\right\}$ on the pupil, with integration performed over the appropriate 
region: 
\begin{equation}
\label{eq:PSF_rect}
I\left( x, y \right) = 
k \left| \int d\xi \, d\eta \, 
P \left( \xi, \eta \right) e^{-i\pi \left( x \xi + y \eta 
\right) } \right| ^{2} \, . 
\end{equation}
As an example, we consider the current Gaia telescope geometry, i.e. 
a rectangular aperture, with size $D_x = 0.5$~m $\times$ $D_y = 1.4$~m 
(high resolution in the $y$ direction) and focal length $F = 35$~m. 
The PSF is described by the $sinc$ squared function, where 
$\rmn{sinc} \, z = \left(\rmn{sin} \, \pi z \right) \, / \, 
\left( \pi z \right)$ \citep{bracewell}: 
\begin{equation} 
\label{eq:PSF_NA}
I\left( x, y \right) = 
\rmn{sinc}^2 \left({x D_x \over \lambda F} \right) \times  
\rmn{sinc}^2 \left({y D_y \over \lambda F} \right) \, . 
\end{equation}
A representation of the non aberrated PSF, at $\lambda = 600$~nm, is shown 
in Fig. \ref{fig_PSF_Gaia}, in logarithmic units normalised to the central 
peak intensity. 

\section{ FT resolution and processing }
\label{app_FT} 
Usually, a uniform sampling period $T$ is assumed for DFT, although this 
is not strictly implicit in the definition \ref{eq:fourier}. 
The signal transform is computed at pulsation $\omega_n = 2 \pi n / 
(NT)$, where $n = 0 \dots (N-1)$, so the resolution is 
$\Delta \omega = 2 \pi / (NT)$ (or $\Delta \nu = 1/ (NT)$, in terms 
of frequency: $\omega = 2 \pi \nu$). 
The DFT thus associates the resolution in one domain to the full range 
of the corresponding conjugate variable. 

The computational load of FFT and DFT of a sequence of $N$ values is 
respectively of order of $N \log N$ and $N^2$, so that for increasing 
$N$ the FFT is much more convenient of DFT performing the same 
computation. 
The expression is valid only when the number of samples is a power of two: 
$N = 2^n$, for some integer $n$; for other values, the result is highly 
dependent on the selected algorithm. 

The numerical implementation of the DFT for the rectangular pupil case 
follows from the definition, apart details of proper definition of the 
pupil and FP coordinates, respectively $\left\{ x_n, \, y_m \right\}$ 
and $\left\{ \xi_p, \, \eta_q \right\}$, with suitable indices 
$\left\{ n, \, m, \, p, \, q \right\}$: 

\begin{equation}
\label{eq:PSF_rect_discr}
I_{nm = }I\left( x_n, y_m \right) = 
k \left| \sum_{p,q} \, 
P \left( \xi_p, \eta_q \right) e^{-i\pi \left( x_n \xi_p + y_m \eta_q 
\right) } \right| ^{2} \, . 
\end{equation}

\label{lastpage}

\end{document}